\documentclass[runningheads]{llncs}
\usepackage{graphicx}

\usepackage{booktabs}  
\usepackage{pbox}
\usepackage{subfigure}
\usepackage{epsfig,endnotes}
\usepackage{epstopdf}
\usepackage{xcolor}
\usepackage{graphicx}
\usepackage{xspace}
\usepackage{enumitem}
\usepackage{url}
\usepackage{flushend}  
\usepackage{arydshln}
\usepackage[ruled,vlined,linesnumbered]{algorithm2e}
\usepackage{amssymb}

\newcommand{\eg}{{e.g.,}\xspace}
\newcommand{\ie}{{i.e.,}\xspace}

\newcommand{\folder}{./}

\smallskip
\smallskip

\newcommand\co[1]{}

\newcommand{\tool}{{BatteryLab}\xspace}
\newcommand{\TOOL}{{BATTERYLAB}\xspace}
\newcommand{\wifi}{{WiFi}\xspace}
\newcommand{\webtool}{{WPM}\xspace}

\newif\ifcomment
\commentfalse

\ifcomment
    \newcounter{MVNumberOfComments}
    \stepcounter{MVNumberOfComments}
    \newcommand{\mvnote}[1]{\textcolor{blue}{\small \bf [MV\#\arabic{MVNumberOfComments}\stepcounter{MVNumberOfComments}: #1]}}
    
    \newcounter{MKNumberOfComments}
    \stepcounter{MKNumberOfComments}
    \newcommand{\mknote}[1]{\textcolor{orange}{\small \bf [MK\#\arabic{MKNumberOfComments}\stepcounter{MKNumberOfComments}: #1]}}
    
\else
    \newcommand\mvnote[1]{}
    \newcommand\mknote[1]{}
\fi

\begin{document}
\title{BatteryLab: A Collaborative Platform for Power Monitoring}
\subtitle{\url{https://batterylab.dev}}
\author{Matteo Varvello\inst{1} \and Kleomenis Katevas\inst{2} \and Mihai Plesa\inst{3} \and Hamed Haddadi\inst{3}\and Fabian~Bustamante\inst{4} \and  Ben Livshits\inst{5}}
\authorrunning{M. Varvello et al.}

\institute{Bell Labs Nokia, \email{matteo.varvello@nokia.com} \and Telefonica Research, \email{kleomenis.katevas@telefonica.com} \and Brave Software, \email{\{mplesa,hhaddadi\}@brave.com} \and Northwestern University, \email{fabianb@cs.northwestern.edu} \and Imperial College London, \email{b.livshits@imperial.ac.uk}}
\maketitle

\begin{abstract}
Advances in cloud computing have simplified the way that both software  development and testing are performed. This is not true for battery testing for which state of the art test-beds simply consist of one phone attached to a power meter. These test-beds have limited resources, access, and are overall hard to maintain; for these reasons, they often sit idle with no experiment to run. In this paper, we propose to \emph{share} existing battery testbeds and transform them into  \emph{vantage points} of \tool, a power monitoring platform offering heterogeneous devices and testing conditions. We have achieved this vision with a combination of hardware and software which allow to augment existing battery test-beds with remote capabilities. \tool currently counts three vantage points, one in Europe and two in the US, hosting three Android devices and one iPhone 7. We benchmark \tool with respect to the accuracy of its battery readings, system performance, and platform heterogeneity. Next, we demonstrate how measurements can be run atop of \tool by developing the ``Web Power Monitor'' (WPM), a tool which can measure website power consumption at scale. We released WPM and used it to report on the energy consumption of Alexa's top 1,000 websites across 3 locations and 4 devices (both Android and iOS). 
\keywords{Battery \and Test-bed \and Performance \and Android \and iOS.}
\end{abstract}

\section{Introduction}
\label{sec:intro}
Power consumption is a growing concern in the mobile industry, ranging from mobile phone users, operating system vendors, and app developers.  To accurately measure a device power consumption, two options are currently available: \emph{software-based} measurements, which rely on battery readings from the device, and \emph{hardware-based} measurements which leverage an external power monitor connected to a device battery. Software-based power measurements are easy to use, but lack the accuracy and granularity an experimenter might require~\cite{schulman2011phone,chenSIGMETRICS15}. Few startups~\cite{greenspector,mobileenerlytics} offer, for a price, improvements upon the accuracy of software-based power measurements by relying on few devices for which they have performed heavy ``calibration'' (their secret sauce). Hardware-based power measurements are accurate, fine-grained, but quite cumbersome to setup. 

For years, researchers have been building home-grown test-beds for hardware-based power measurements, consisting of an Android device connected to a high-frequency power monitor~\cite{buiMOBICOM15,caoPOMAC17,ravenMOBICOM17,thiagarajanWWW12}. This required expertise in hardware setup and writing code when automation is needed -- code which is unfortunately never shared with the community.  Such closed-source test-beds have limited accessibility, \eg requiring physical access to the devices, and shareability, even among members of the same group. This became clear during the COVID-19 pandemic: remote desktop tools like VNC came to the rescue, but often the only solution was to move that precious test-bed at home. 

\begin{figure}[t]
    \centering
    \psfig{figure=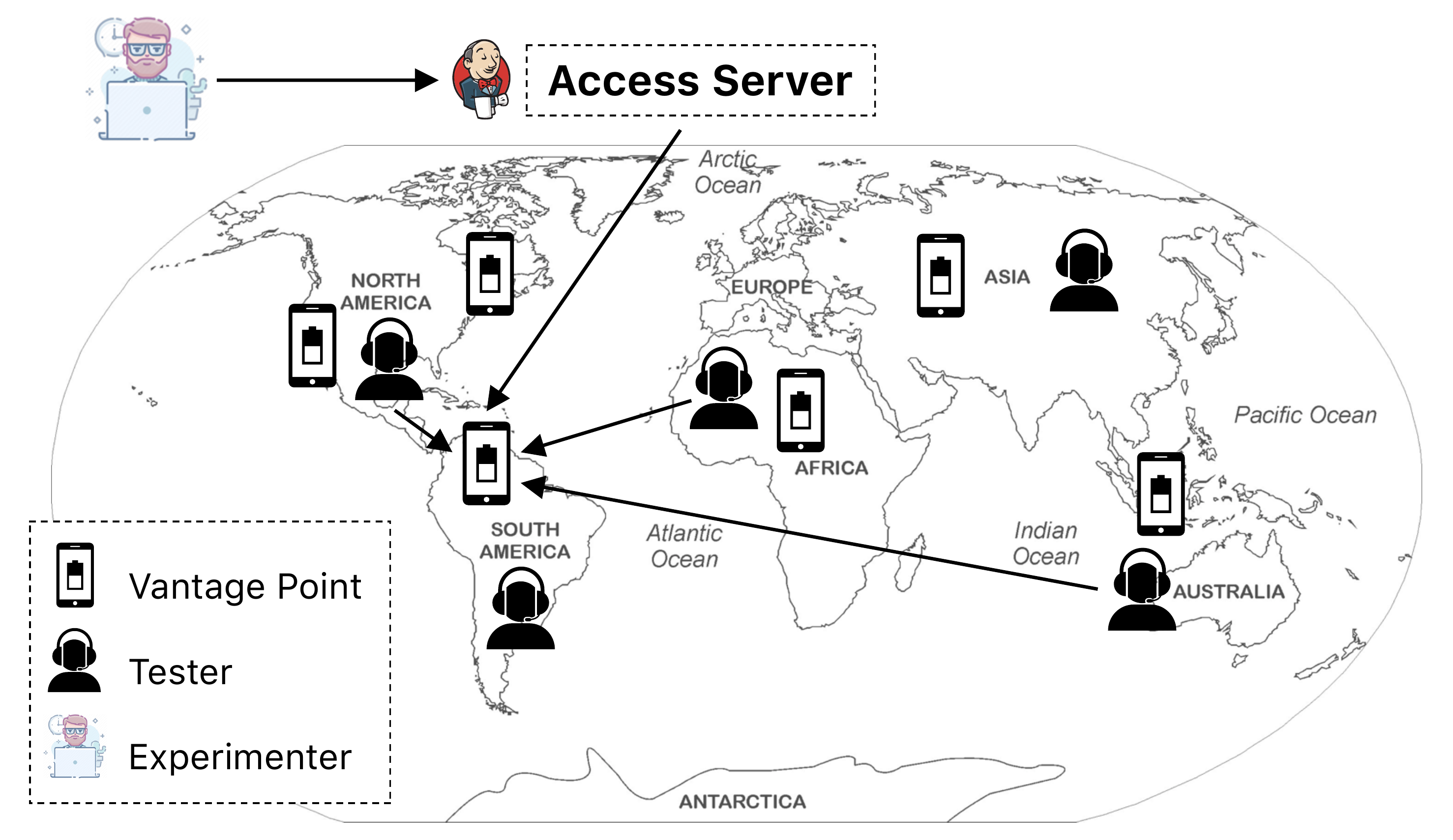, width=0.9\columnwidth}
    \caption{Distributed architecture of \tool.}
    \label{fig:arch-a}
\end{figure}

In this paper  we challenge the assumption that such battery test-beds need to be ``local'' and propose \emph{\tool}, a cooperative platform for battery measurements. We envision \tool as a cooperative platform where members contribute hardware resources (\eg~some phones and a power monitor) in exchange for access to the resources contributed by other platform members. Nevertheless, the hardware/software suite we have built and open sourced~\cite{our-code} can also be used ``locally'', \ie augmenting an existing battery test-bed with scheduling and remote control capabilities. The following contributions are the founding blocks of \tool:

\vspace{0.1in}
\noindent\textbf{Automation for hardware-based power measurements.} \tool comes with an intrinsic automation requirement. For example, an \emph{experimenter} from Europe needs to be able to activate a power meter connected to a phone in the US. To make this possible, we have designed \emph{vantage points} as the above local test-beds enhanced with a lightweight \emph{controller} such as a Raspberry Pi~\cite{rasbpi}. The controller runs \tool's software suite which realizes ``remote power testing'', \eg~from activating a device's battery bypass to enabling remote control of the device via the experimenter's browser. 

\vspace{0.1in}
\noindent\textbf{A library for Android and iOS automation.} While the Android Debugging Bridge (ADB) is a powerful tool to automate Android devices, an equivalent does not exist for iOS. \tool builds atop of ADB to offer seamless automation of Android devices. For iOS, we have built and open-sourced a Python library which maps commands like touch, swipe, and text input to a (virtual) Bluetooth keyboard and mouse. To the best of our knowledge, we are the first to provide automation of any third party app on actual iOS devices (\ie~other than simulators as in~\cite{appetize,runthatapp}). Even commercial products for iOS, such as TeamViewer~\cite{teamviewer} or the recent SharePlay~\cite{shareplay} of iOS~15, can only provide remote screen sharing. 

\vspace{0.1in}
\noindent\textbf{Usability testing for power measurements.} \tool allows an experimenter to interact with a real device via its browser. This feature is paramount for debugging automation scripts, but also a key enabler of \emph{usability testing}, or battery measurements coupled with actual device interactions from real users. 

\vspace{0.1in}
\noindent\textbf{Deployment at three research institutions.} \tool currently has three vantage points, two in the US and one in Europe (with more vantage points going live soon) and hosts a range of Android devices and an iOS device (iPhone~7). 

\vspace{0.05in}
We evaluate \tool on battery readings accuracy, system performance, and platform heterogeneity. To illustrate the value and ease-of-use of \tool, we have also built the ``Web Power Monitor'' (\webtool), a service which measures the power consumption of  websites loaded via a test browser running at any \tool's device. With a handful of lines of code, \webtool allowed us to conduct the largest scale measurement study of energy consumption on the Web, encompassing Alexa's top 1,000 websites measured from four devices and two operating systems. We have released \webtool as a web application integrated with \tool which offers such testing capabilities to the public, in real time. 
This paper extends our previously published work~\cite{varvello2019batterylab} in many ways: 

\begin{itemize}
  \item We add support for device automation also to Apple iOS by exploiting the Bluetooth HID and AirPlay services.
  \item We deploy \tool at three research institutions and benchmark its performance including, among others, a comparison with software-based battery measurements. 
  \item We open source \tool's code for ``local'' use, and \tool as a testbed for battery measurements. 
  \item We develop and release \webtool, a tool for measuring website power consumption at scale; we further use \webtool to measure the energy consumption of Alexa's top 1,000 websites across 3 locations and 4 devices.
   \item We explore support for usability testing via ``action replay'', a mechanism to automatically build app automation scripts based on human inputs. 
\end{itemize}
\section{\TOOL Architecture}
\label{sec:system}
This section presents the design and implementation details of \tool (see Figure~\ref{fig:arch-a}). Our current iteration focuses on mobile devices, but the architecture is flexible and can be extended to other devices, \eg~laptops and IoT devices.

\tool consists of a centralized \emph{access server} that remotely manages a number of nodes or \emph{vantage points}. Each of these vantage points, hosted by universities or research organizations around the world, includes a number of test devices (a phone/tablet connected to a power monitor) where experiments are carried out. \tool members (\emph{experimenters}) gain access to test devices via the access server, where they can request time slots to deploy automated scripts and/or remote control of the device. Once granted, remote device control can be shared with \emph{testers}, whose task is to manually interact with a device, \eg~scroll and search for items on a shopping application. Testers are either volunteers, \eg~recruited via email or social media, or paid, recruited via crowdsourcing websites like Mechanical Turk~\cite{mturk}. 

In the remainder of this section, we describe \tool's  main components in detail. Next, we focus on \tool's automation capabilities and on the procedure for new members to join the platform. 

\subsection{Access Server}
\label{sec:sys:server}
The main role of the access server is to manage the vantage points and  schedule experiments on them based on experimenters' requests. We built the access server atop of the Jenkins~\cite{jenkins} continuous integration system which is free, open-source, portable (written in Java) and backed by an active and large community. Jenkins enables end-to-end test pipelines while supporting multiple users and concurrent timed sessions.

\tool's access server runs in the cloud (Amazon AWS) which enables further scaling and cost optimization. Vantage points have to be added explicitly and pre-approved in multiple ways (IP lockdown, security groups). Experimenters need to authenticate and be authorized to access the web console of the access server, which is only available over HTTPS. The access server communicates with the vantage points via SSH. New \tool members grant SSH access from the server to the vantage point's controller via public key and IP white-listing (\S~\ref{sec:sys:join}). 

Experimenters access vantage points via the access server, where they can create \emph{jobs} to automate their tests. Jobs are programmed using a combination of \tool's Python API (Table~\ref{tab:api}), \eg~for user-friendly device selection and interaction with the power meter, and code specific to each test. Only the experimenters who have been granted access to the platform can create, edit, or run jobs and every pipeline change has to be approved by an administrator. This is done via a role-based authorization matrix.

After the initial setup, the access server dispatches queued jobs based on the experimenter constraints, \eg~target device, connectivity, or network location, and \tool constraints. For example, no concurrent jobs are allowed at the same vantage point since the power monitor can only be associated with one device at a time and isolation is required for accurate power measurements. By default, the access server collects logs from the power meter which are made available for several days within the job's workspace. Android logs (\eg~\texttt{logcat} and \texttt{dumpsys}) can be requested via the \texttt{execute\_command}  API for the supported devices (Table~\ref{tab:api}).

\begin{figure}[t]
    \centering
    \psfig{figure=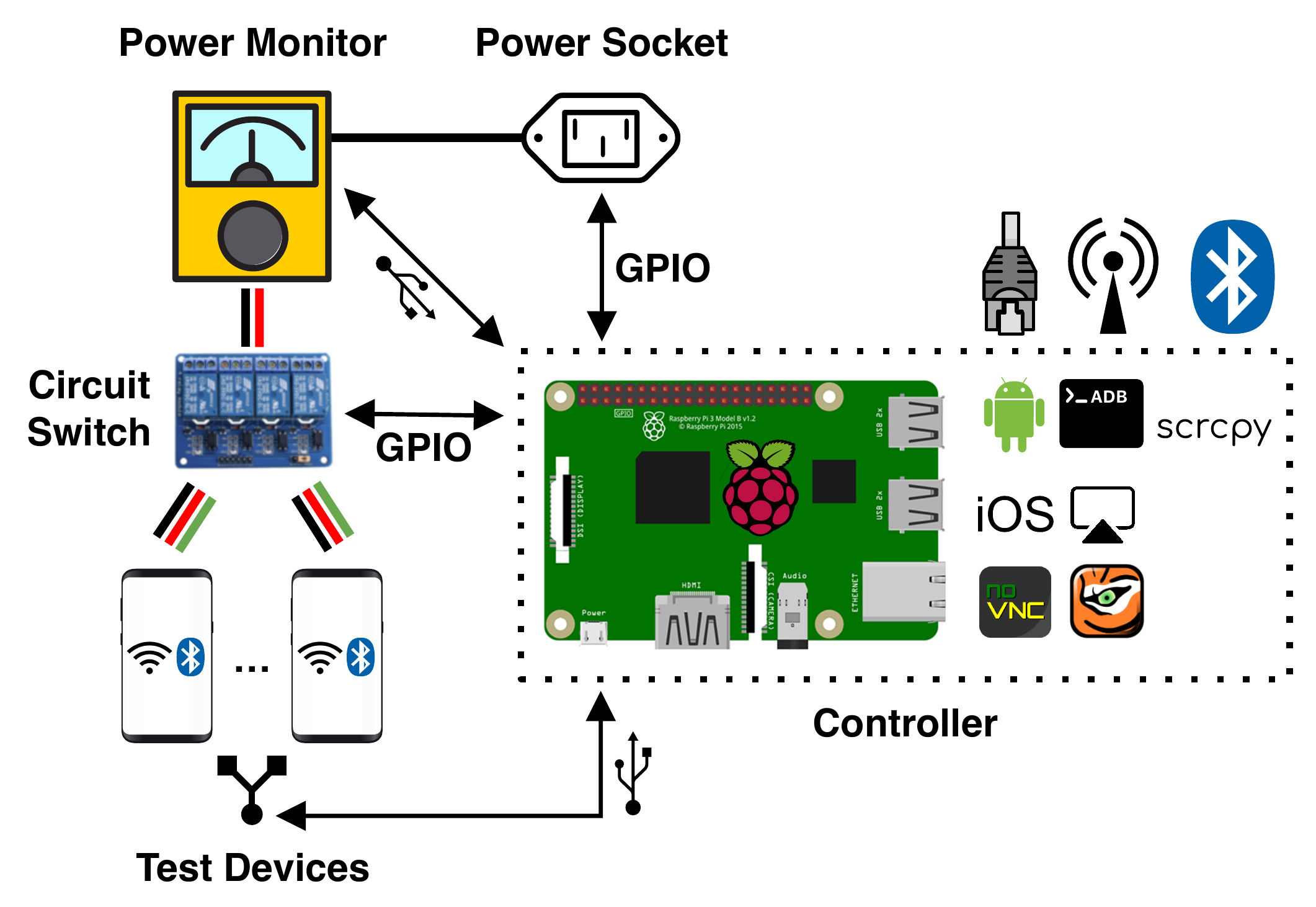, width=0.7\columnwidth}
    \caption{Vantage point design.}
    \label{fig:arch-b}
\end{figure}

\subsection{Vantage Point}
\label{sec:sys:remote}
Figure~\ref{fig:arch-b} shows a graphical overview of a \tool's vantage point with its main components: controller, power monitor, test devices, circuit switch, and power socket. 

\vspace{0.1in}
\noindent\textbf{Controller} -- This is a Linux-based machine responsible for managing the vantage point. This  machine is equipped with both Ethernet, \wifi and Bluetooth connectivity, a USB controller with a series of available USB ports, as well as with an external General-Purpose Input/Output (GPIO) interface. We use the popular Raspberry Pi 3B+~\cite{rasbpi} running Raspberry~Pi OS (Buster, September 2019) that meets these requirements at an affordable price.

The controller's primary role is to manage connectivity with test devices. Each device connects to the controller's USB port, \wifi access point (configured in NAT or Bridge mode), and Bluetooth, based on automation needs (see Section~\ref{sec:sys:automation}). USB  is used to power each testing device when not connected to the power monitor and to instrument Android devices via the Android Debugging Bridge~\cite{adb} (ADB), when needed. \wifi provides Internet access to all devices and extend ADB automation and device mirroring to Android devices without incurring the extra USB current, which interferes with the power monitoring procedure. (De)activation of USB ports is realized using \texttt{uhubctl}~\cite{uhubctl}. Bluetooth is used to realize automation across OSes (Android and iOS) and connectivity (\wifi and cellular). 

The second role of the controller is to provide \emph{device mirroring}, \ie~remote control of device under test. We use VNC (\texttt{tigervnc}~\cite{tigervnc}) to enable remote access to the controller, and \texttt{noVNC}~\cite{novnc}, an HTML VNC library and application, to provide easy access to a VNC session via a browser without additional software required at the experimenter/tester. We then \emph{mirror} the test device within the noVNC/VNC session  and limit access to only this visual element.  In Android, this is achieved using \texttt{scrcpy}~\cite{scrcpy}, a screen mirroring utility which runs atop of ADB for devices running API 21 (Android $\geq$ 5.0). In iOS, we utilize AirPlay Screen Mirroring~\cite{AirPlay} using \texttt{RPiPlay}~\cite{RPiPlay}, an AirPlay mirroring server for devices running iOS $\geq$ 9.0. 

We have also built a graphical user interface (GUI) around the default \texttt{noVNC} client. The GUI consists of an \emph{interactive area} where a device screen is mirrored (bottom of the figure) while a user (experimenter or tester) can remotely mouse-control the physical device, and a \emph{toolbar} that occupies the top part of the GUI and implements a convenient subset of \tool's API (see Table~\ref{tab:api}). 

\vspace{0.1in}
\noindent\textbf{Power Monitor} -- This is a power metering hardware capable of measuring the current consumed by a test device in high sampling rate. \tool currently supports the Monsoon HV~\cite{monsoon}, a power monitor with a voltage range of 0.8V to 13.5V and up to 6A continuous current sampled at 5~KHz. The Monsoon HV is controlled using its Python API~\cite{pymonsoon}. Other power monitors can be supported, granted that they offer APIs to be integrated with \tool's software suite.

\vspace{0.1in}
\noindent\textbf{Test Device(s)} -- It is an Android or iOS device (phone or tablet) that can be connected to a power monitor using a battery bypass modification (\ie~isolate the battery power circuit and provide power via the power monitor). While devices with removable batteries are easier to setup, more complex configurations (\eg~all iOS and recent Android devices) are also supported by doing the battery bypass modification at the battery controller level.

\vspace{0.1in}
\noindent\textbf{Circuit Switch} -- This is a relay-based circuit with multiple channels that lies between the test devices and the power monitor. The circuit switch is connected to the controller's GPIO interface and all relays can be controlled via software from the controller. Each relay uses the device's voltage (+) terminal as an input, and programmatically switches between the battery's voltage terminal and the power monitor's \texttt{Vout} connector. Ground (-) connector is permanently connected to all devices' Ground terminals. 

This circuit switch has three main tasks. First, it allows to switch between a direct connection between the phone and its battery, and the ``battery bypass''---which implies disconnecting the battery and connecting to the power monitor. This is required to allow the power monitor to measure the current consumed during an experiment. Second, it allows \tool to concurrently support multiple test devices without having to manually move cables around. Third, it allows to programmatically switch the power meter on and off. 

\vspace{0.1in}
\noindent\textbf{Power Socket} -- This is a relay-based power socket that allows the controller to turn the Monsoon on and off, when needed. It connects to the controller via the GPIO port, and it is controlled by our Python API.
\begin{table}[t]
\small
\centering
\begin{tabular}{rccc}
\hline
    {\bf API}     & {\bf Description}          & {\bf Parameters}  \\
    \hline
    {\bf list\_nodes}       & \begin{tabular}{@{}c@{}}List matching \\ vantage points\end{tabular}  & label, state  \\
    \hdashline[0.5pt/5pt]
    {\bf list\_devices}       & \begin{tabular}{@{}c@{}}List identifiers \\ of test devices\end{tabular}  & vantage\_point  \\
    \hdashline[0.5pt/5pt]
    {\bf device\_mirroring}   & \begin{tabular}{@{}c@{}}Activate device \\ mirroring\end{tabular}    & device\_id    \\
    \hdashline[0.5pt/5pt]
    {\bf power\_monitor}      & \begin{tabular}{@{}c@{}}Toggle Monsoon\\ power state\end{tabular}   & state (on/off)             \\
    \hdashline[0.5pt/5pt]
    {\bf set\_voltage}        & \begin{tabular}{@{}c@{}}Set target\\ voltage \end{tabular}          & voltage\_val  \\
    \hdashline[0.5pt/5pt]
    {\bf start\_monitor}      & \begin{tabular}{@{}c@{}}Start battery\\ measurement \end{tabular}    & device\_id, duration\\
    \hdashline[0.5pt/5pt]
    {\bf stop\_monitor}       & \begin{tabular}{@{}c@{}}Stop battery\\ measurement \end{tabular}     & -             \\
    \hdashline[0.5pt/5pt]
    {\bf batt\_switch}        & \begin{tabular}{@{}c@{}}(De)activate\\ battery \end{tabular}         & device\_id    \\
    \hdashline[0.5pt/5pt]
    {\bf execute\_command}        & \begin{tabular}{@{}c@{}}Execute a \\ command on device \end{tabular}   &  \begin{tabular}{@{}c@{}}device\_id, command,\\ automation \end{tabular} \\
    \hdashline[0.5pt/5pt]
\end{tabular}
\caption{\tool's core API.}
\label{tab:api}
\end{table}

\section{Using BatteryLab}
\label{sec:system_usage}
In the following paragraphs we illustrate the use of \tool's API, discuss its support of test automation, and the generation of automation scripts from human input. We close the section with a description of the steps needed to join \tool.

\subsection{API Usage}
\label{sec:sys:api}
Experimenter jobs are interleaved with ``control'' jobs which manage the vantage points, \eg~they update \tool wildcard certificates (\S~\ref{sec:sys:join}) and ensure that the power meter is not active when not needed (for safety reasons). We here present some of these jobs as examples of \tool's API usage. We have chosen the set of jobs that are also used by the application we have built atop of \tool (Section~\ref{sec:webtest}). Note that these jobs effectively \emph{extend} the API available to \tool's experimenters; these are not listed in Table~\ref{tab:api} which focuses only on core API.

\vspace{0.1in}
\noindent \texttt{NODE\_SETUP} -- The goal of this job/API is to prepare a vantage point for  power measurements on a device $d$. This implies activating the power meter (\texttt{power\_monitor}), offering the voltage that $d$ requires (\texttt{set\_voltage}) and activating the relay to realize $d$'s battery bypass (\texttt{batt\_switch}). The job continues by verifying that \wifi is properly working, eventually switching frequency based on the device characteristics --- with 5~GHz preferred, when available. Based on the device and the requested automation (see Section~\ref{sec:sys:automation}), the job continues by either activating ADB over \wifi or the Bluetooth HID service. Finally, USB connection is interrupted --- to avoid noise on the power measurements --- and device mirroring is activated, if needed (\texttt{device\_mirroring}).

\vspace{0.1in}
\noindent \texttt{DEVICE\_SETUP} -- The goal of this job/API is to prepare a device $d$ such that ``noise'' on the upcoming power measurement is minimized. We have identified several best practices which help in doing so and we offer them as an API. Nevertheless, the experimenter is the ultimate decision maker and can either ignore or further improve on these operations. The job starts by disabling notifications, set the device in airplane mode with \wifi only activated --- unless a mobile connection is needed and available --- and close all background apps. Next, the job ensures that the device is not using automatic brightness and further sets the brightness to a default value or a requested one. The last step is important since the variation in ambient light can impact the outcome of a measurement.

\vspace{0.1in}
\noindent \texttt{CLEANUP} -- The goal of this job/API is to ensure that a vantage point is in a ``safe'' state. This implies turning off the power meter if no testing job is undergoing and removing any eventual battery bypass. Finally, USB connectivity is re-enabled which ensures that the device's battery get charged. This job further proceeds removing installed apps which were not used in the last seven days, with the goal to avoid overloading testing devices. 

\vspace{0.1in}
\noindent \texttt{REFRESH} -- The goal of this job/API is to verify reachability of vantage points and devices therein. The information collected is used to populate a JSON file which enhances Jenkins data past sites reachability via SSH. This job currently runs across the whole platform every 30 minutes. 

\subsection{Android / iOS Automation Library}
\label{sec:sys:automation}
\tool provides a Python library --- which we open-sourced together with the \tool's code --- that greatly simplifies test automation on both Android and iOS. At high level, the library offers APIs like \texttt{input(tap, x, y)} which map to several underlying automation mechanisms, each with its own set of advantages and limitations. The library automatically switches to an automation solution based on the experiment needs, \eg device and connectivity, hiding unnecessary complexity to the experimenter.

\vspace{0.1in}
\noindent \textbf{Android Debugging Protocol} (Android) --  ADB~\cite{adb} is a powerful tool/protocol to control an Android device. Commands can be sent over USB, \wifi, or Bluetooth. While USB guarantees highest reliability, it interferes with the power monitor due to the power required to activate the USB micro-controller at the device. Accordingly, \tool's automation library uses ADB over USB \emph{whenever} the power monitor is not used, \eg when installing an app or cleaning a device, while resorting to \wifi (or Bluetooth)  for all other automations. Note that using \wifi implies not being able to run experiments leveraging the mobile network. However, these experiments are possible leveraging Bluetooth tethering, when available. 

\vspace{0.1in}
\noindent \textbf{Bluetooth HID Service} (iOS/Android) -- Automating third-party apps in iOS is challenging due to the lack of ADB-like API. Even commercial solutions like TeamViewer~\cite{teamviewer} or the new SharePlay~\cite{shareplay} of iOS~15 limit their iOS offering to remote screen viewing only. The only solution to control an iOS device without physical access requires using a wireless keyboard and mouse. We exploit this feature to map commands like touch, swipe and text input into (virtual) mouse and keyboard actions. 

Specifically, we virtualize the mouse and keyboard by designing a Human Interface Device (HID) service~\cite{bluetooth_hid} atop of BlueZ Bluetooth Protocol Stack~\cite{bluez} v5.43. The controller broadcasts a custom Combo Keyboard/Pointing HID service (\ie~HIDDeviceSubclass: \texttt{0xC0}~\cite{bluetooth_hid}) which enables a connection to previously paired test devices over Bluetooth. The automation library translates keyboard keystrokes, mouse clicks and gestures into USB HID Usage Reports~\cite{usb_hid} that simulate user actions to the controlled device (\eg~locate an app, launch it, and interact with it). While we exploit this automation strategy for iOS only, the approach is generic and can be used across all devices which support the Bluetooth HID profile for both mouse and keyboard (\ie~Android v8.0+ and iOS v13.0+).

\subsection{Action Replay}
\label{sec:sys:action}
Regardless of the automation mechanism used, building automation scripts for mobile devices is a time consuming task~\cite{leung2016appforthat,lucky2015wisec,jingjingrecon}. Device mirroring offers a unique opportunity to speed up the generation of such automation scripts in \tool. The key idea is to record an experimenter/tester clicks, mouse, keyboard input, and use them to generate an automation script. 

We have thus modified noVNC -- precisely \texttt{mouse.js} and \texttt{keyboard.js} -- to POST the collected user input to the controller's web application (see Section~\ref{sec:sys:remote}) where the device being mirrored is hosted. The web application collects the user input and map it to APIs from the above automation library, which translates into, for example, an ADB command such as \texttt{tap} or \texttt{swipe}. When screen coordinates are involved, \eg~for a \texttt{tap} command, the actual coordinates are derived by offsetting the coordinates recorded in noVNC as a function of the size of the VNC screen and the actual device size. Under the assumption that an application GUI is similar across platforms, the human-generated automation script at a given device could be re-used for other devices. 

 \subsection{How to Join?}
 \label{sec:sys:join}
Joining \tool is straightforward and consists of three steps. First, the vantage point needs to be physically built as described in Figure~\ref{fig:arch-b}. At this point, the controller (Raspberry Pi) should also be flashed with the latest Raspberry~Pi OS image along with some standard setup as described in the associated tutorial~\cite{blabtutorial}.  Second, the network where the controller is connected (via Ethernet) needs to be configured to allow the controller to be reachable at the following configurable ports: 2222 (SSH, access server only)\footnote{The SSH agent at the node also needs to be configured accordingly. An iptable rule should be added to limit access to the the access server only.}, 8080 (web application for GUI and action/replay).  Third, a \tool account should be created for the new member. This involves downloading the access server's public key --- to be authorized at the controller --- and uploading a human readable identifier for the vantage point (\eg~\texttt{node1}), and its current public IP address. This information is used by the access server to add a new entry in \tool's DNS (\eg~\texttt{node1.batterylab.dev}) --- provided by Amazon Route53~\cite{route53} --- and verify that SSH access to the new vantage point is now granted. Since the whole \tool traffic is encrypted, a wildcard \texttt{letsencrypt}~\cite{letsencrypt} certificate is distributed to new members by the access server, which also manages its renewal and distribution, when needed. 

The next step consists of installing \tool's software at the controller. This step is realized automatically by \tool's access server and it is the first job to be deployed at the new vantage point. At high level, this consists in the following operations. First, the OS is updated. Next, common security practices are enforced: 1) install \texttt{fail2ban} which neutralizes popular brute-force attacks over SSH, and 2) disable password authentication for  SSH. Next, \tool code is pulled from its open source repository~\cite{our-code} along with all packages and software needed. Code is compiled, where needed, and packages are installed. Then, the controller is turned into an ``access point'' where the test devices will connect to. By default, the access point spins a new SSID (\texttt{BatteryLab}) with a pre-set password operating on 2.4~GHz. However, \tool automatically switches to 5~GHz for devices that support it. This ``switch'' is required since the Raspberry Pi does not mount two \wifi antennas and thus both frequencies cannot be active at the same time. 

Next, several \texttt{crontab} entries are added. At reboot and every 30 minutes, a task monitors the controller's public IP address and update its entry at \tool's DNS. At reboot, the GPIO pins used by \tool are set as ``output'' and the IP rules needed by the controller to act as an access point are restored. The next step consists in setting up device mirroring, \ie~VNC password and wildcard certificate used by both noVNC and the web application. This setup job also learns useful information about the devices connected: ADB identifier (if available), screen resolution, IP address, etc. This information is reported to the access server to further populate the JSON file maintained by the \texttt{REFRESH} job/API. Last but not least, several tests are run to verify: 1) Monsoon connectivity, 2) device connectivity, 3) circuit relay stability, 4) device mirroring. 

\tool currently counts three vantage points located in the UK, New Jersey, and Illinois, with a total of three Android devices and one iPhone 7. Table~\ref{tab:testbed} provides detailed information of the devices currently available to the public via \tool. At the time of writing, three other organizations are in the process of setting up a \tool vantage point. 
\begin{table*}[t]
\small
\centering
\begin{tabular}{p{2.2cm} p{2cm} p{2.7cm} p{2.7cm} p{2.4cm}}
\hline
                      & {\bf J7DUO} & {\bf IPHONE7} &  {\bf SMJ337A} & {\bf LMX210}\\
     \hline
     {\bf Vendor}     & Samsung & Apple & Samsung   & LG \\ 
     {\bf OS}       & Android 9.0 & iOS 13.2.3  & Android 8.0.0  & Android 7.1.2   \\
     {\bf Location} & United Kingdom & United Kingdom &  New Jersey & Illinois \\
     {\bf CPU Info} & Octa-core (2x2.2 GHz Cortex-A73, 6x1.6 GHz & Quad-core 2.34 GHz Apple A10 Fusion &  Quad-core 1.4 GHz Cortex-A53 & Quad-core 1.4 GHz Cortex-A53\\
     {\bf Memory}   & 4GB & 2GB & 2GB & 2GB\\
     {\bf Battery}  & 3,000mAh &  1,960mAh & 2,600mAh & 2,500mAh\\
\hline
\end{tabular}
\caption{\tool test-bed composition.}
\label{tab:testbed}
\end{table*}
\section{Benchmarking}
\label{sec:bench}
This section benchmarks \tool. We first evaluate its \emph{accuracy} in reporting battery measurements. We then evaluate its \emph{performance} with respect to CPU, memory, and responsiveness of its device mirroring mechanism. We then investigate \tool's \emph{heterogeneity} and the feasibility of usability testing when coupled with power monitoring.  

\subsection{Accuracy}
\label{sec:eval:accuracy}
Compared to a classic \emph{local} setup for device performance measurements, \tool introduces some hardware (circuit relay) and software (device mirroring) components that can impact the \emph{accuracy} of the measurements. We devised an experiment where we compare three scenarios. First, a \emph{direct} scenario consisting of the Monsoon power meter, the testing device, and the Raspberry Pi to instrument the power meter. For this setup, we strictly followed Monsoon indications~\cite{monsoon} in terms of cable type and length, and connectors to be used. Next we evaluate a \emph{relay} scenario, where the relay circuit is introduced to enable \tool's programmable switching between battery bypass and regular battery operation (see Section~\ref{sec:sys:remote}). Finally, a \emph{mirroring} scenario where the device screen is mirrored to an open noVNC session. While the relay is always ``required'' for \tool to properly function, device mirroring is only required for usability testing. Since we currently do not fully support usability testing for iOS (see Section~\ref{sec:sys:server}), we here only focus on Android. 

\begin{figure}[t]
    \centering
    \psfig{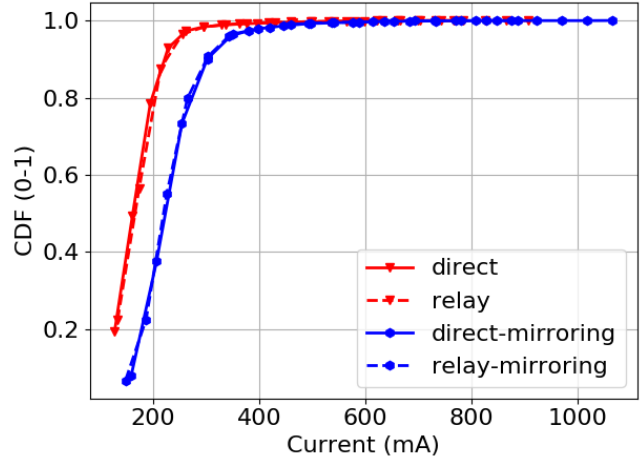}
    \caption{CDF of current drawn(direct,  relay,  direct-mirroring, relay-mirroring).}
    \label{fig:relay}
\end{figure}

Figure~\ref{fig:relay} shows the Cumulative Distribution Function (CDF) of the current consumed in each of the above scenarios during a 5 minutes test. For completeness, we also consider a \emph{direct-mirroring} scenario where the device is directly connected to Monsoon and device mirroring  is active. During the test, we play an MPEG4 video pre-loaded on the SD card of the device (J7DUO, UK). The rationale is to force the device mirroring mechanism to constantly update as new frames are initiated. The figure shows negligible difference between the ``direct'' and ``relay'' scenarios, regardless of the device mirroring status being active or not. A larger gap (median current grows from 160 to 220~mA) appears with device mirroring. This is because of the background process responsible for screencasting to the controller which causes additional CPU usage on the device ($\sim$15\%). At the end of this section, we investigate a more challenging usability testing scenario along with a potential solution to minimize the additional power consumption caused by device mirroring. 

\begin{figure}[t]
    \centering
    \psfig{figure=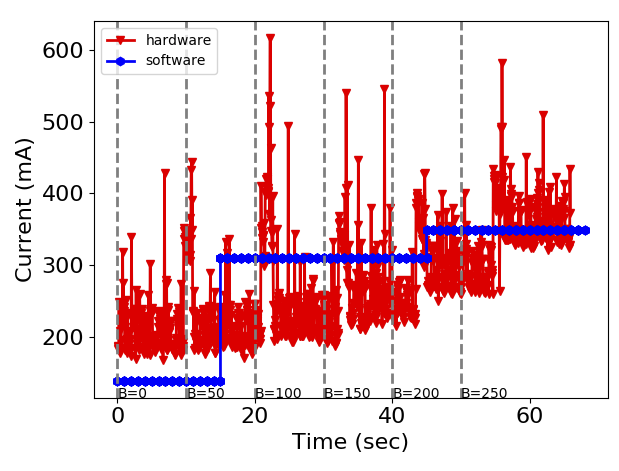, width=0.7\columnwidth}
    \caption{Current over time under variable screen brightness ($B$) from 0 to 250, Android's max value. Software versus hardware measurements (LMX210).}
    \label{fig:res-soft-hard}
 \end{figure}

A related question is: \emph{what is the accuracy that \tool offers compared to software measurements?} Having verified that \tool is as accurate as a local setup, and granted that hardware-based battery measurements are the ``ground truth'', the question is really how accurate are software-based battery measurements? While this question is out of scope for this paper, it has to be noted that  Android software-based battery readings can be realized in \tool via ADB\footnote{Either using Android bug-report files or with \texttt{adb shell cat sys/class/power\_supply/*/uevent}}. With respect to iOS, while some  high level \emph{energy usage} reports are available ---  reporting battery consumption every second on an arbitrary  0 to 20 scale --- they are currently unavailable to \tool since they require a developer-enabled macOS.

We find that pure software measurements are \emph{enough} to identify trends in measured current, but have limited overall accuracy and granularity, \eg~a 30 seconds reporting frequency across all our Android devices. As an example, Figure~\ref{fig:res-soft-hard} shows the time evolution of the current measured via \tool (hardware) and software while increasing the screen brightness from minimum (0) to maximum (250) by 50 units over 60 seconds, as indicated by the vertical dashed lines. This plot shows that pure software measurements are \emph{enough} to identify trends, but have limited overall accuracy and granularity -- while the plot refers to the LMX210, we measured a similar reporting frequency (30 seconds) across all Android devices. 

To further investigate the reporting frequency, we have performed the same test also on Samsung's Remote Test Lab~\cite{samsung}.\footnote{These tests were not possible on AWS Device Farm~\cite{awsfarm} due to lack of ADB access.}  We find a 10 seconds reporting frequency on Samsung Galaxy S5 (Android 6) and S7 (Android 8), and 30 seconds on S8 and S9 (Android 9). When repeating the same tests on newer models, we find that the reported sampling rate improves to a mean of $2.23$ sec. ($\pm 1.65$) for Google Pixel 3a (Android 12), $0.66$ ($\pm 0.24$) for Google Pixel 4 (Android 12) and $0.60$ ($\pm 0.25$) for Google Pixel 5. The sampling rate was unaffected from different configurations (screen on, off, or streaming a HD video). Note that internal battery readings can be enhanced with additional data (\eg~cpu, screen usage), alongside device calibration, to achieve higher accuracy, as discussed in~\cite{chenSIGMETRICS15}.

\subsection{System Performance}
\label{sec:eval:perf}
\begin{figure}[t]
    \centering
    \psfig{figure=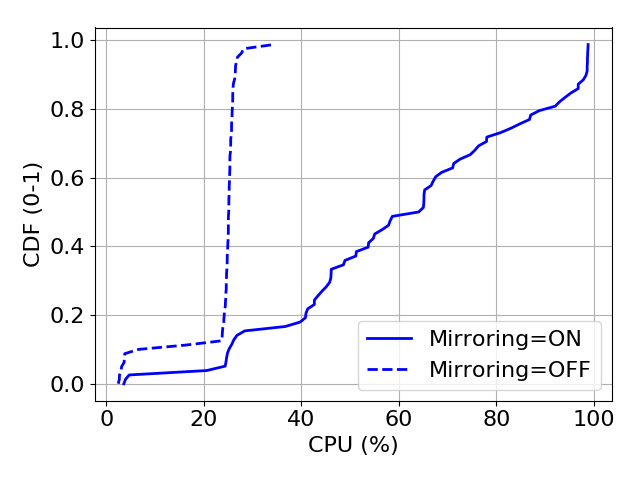, width=0.7\columnwidth}
    \caption{CDF of CPU consumption at the controller (Raspberry Pi 3B+)}
    \label{fig:res:cpu}
\end{figure}

Next, we benchmark overall \tool performance. We start by evaluating the CPU utilization at the controller. Figure~\ref{fig:res:cpu} shows the CDF of the CPU utilization during the previous experiments (when a relay was used) with active and inactive device mirroring, respectively. When device mirroring is inactive, the controller is mostly underloaded, \ie~constant CPU utilization at 25\%. This load is caused by the communication with the power meter to pull battery readings at the highest frequency (5~kHz). With device mirroring, the median load increases to $\sim$75\%. Further, in 10\% of the measurements the load is quite high and over 95\%. 

Device mirroring only impacts the CPU usage. The impact on memory consumption is minimal (extra 6\%, on average). Overall, memory does not appear to be an issue given less than 20\% utilization of the Raspberry Pi's 1~GB. The networking demand is also minimal, with just 32~MB of upload traffic for a $\sim$7 minutes test (due to device mirroring). Note that we set \texttt{scrcpy}'s video encoding (H.264) rate to 1~Mbps, which produces an upper bound of about 50~MB. The lower value depends on extra compression provided by \texttt{noVNC}. 

Finally, we investigate the ``responsiveness'' of device mirroring. We call \emph{latency} the time between when an action is requested (either via automation or a click in the browser), and when the consequence of this action is displayed back in the browser, after being executed on the device. This depends on a number of factors like network latency (between the browser and the test device), the load on the device and/or the controller, and software optimizations. We estimate such latency by recording audio (44,100~Hz) and video (60~fps) while interacting with the device via the browser. We then manually annotated the video using ELAN multimedia annotator software~\cite{elan} and compute the latency as the time between a mouse click (identified via sound) and the first frame with a visual change in the app. We repeat this test 40 times while co-located with the vantage point (1~ms network latency) and measure an average latency of 350 ($\pm$80)~ms.

\begin{figure}[t]
    \centering
    \psfig{figure=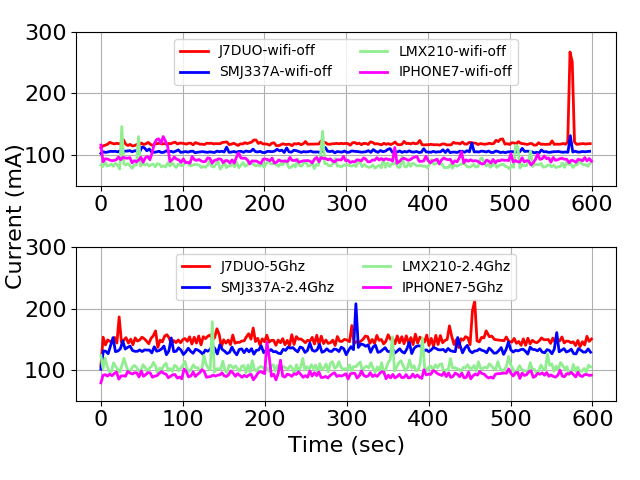, width=0.95\columnwidth}
    \caption{Time evolution of current usage per device at rest (\wifi off and on).}
    \label{fig:current-time}
\end{figure}

\subsection{Devices and Locations}
\label{sec:eval:vpn}
\tool's distributed nature is both a \emph{feature} and a \emph{necessity}. It is a feature since it allows battery measurements under diverse device and network conditions which is, to the best of our knowledge, a first for research and development in this space. It is a necessity since it is the way in which the platform can scale without incurring high costs. We here explore the impact of such diversity on battery measurements. 

Figure~\ref{fig:current-time} displays the evolution over time (600 seconds) of the current used by each \tool device at ``rest'', \ie~displaying the default phone desktop after having run \tool's API \texttt{DEVICE\_SETUP} (Section~\ref{sec:sys:server}) to ensure equivalent device settings. We further differentiate between the case when \wifi was active or not. For Android, regardless of \wifi settings, the figure shows that the J7DUO consumes the most, while the LMX210 consumes the least -- about 25\% less (270 vs 359~J over 600 seconds). Overall, the similar results in the case without \wifi suggest that the difference between the device is intrinsic of the device configurations, \eg~more power-hungry hardware and different Android versions with potential vendor customization. Understanding the event responsible of the variations shown in Figure~\ref{fig:current-time} is out of the scope of this analysis. It is worth noticing some correlations between peaks suggesting vendor specific operations, \eg~the peak around 580 seconds for the J7DUO and SMJ337A, both Samsung devices. The IPHONE7 consumes the least when considering active \wifi, while the LMX210 consumes the least in absence of \wifi. The take away of this analysis is that \tool's devices (and locations) have the potential to offer a large set of heterogeneous conditions for the experimenters to test with. 

\begin{table}[t]
\small
\centering
\begin{tabular}{rcccc}
\hline
    {\bf Speedtest Server (Kms)}   & {\bf Download (Mbps)} & {\bf Upload (Mbps)} & {\bf Latency (ms)}\\
\hline
    \begin{tabular}{@{}c@{}}South Africa\\ Johannesburg (3.21)\end{tabular}  & 6.26  & 9.77  & 222.04 \\
    \hdashline[0.5pt/5pt]
    \begin{tabular}{@{}c@{}}China\\  Hong Kong (4.86)\end{tabular}           & 7.64  & 7.77  & 286.32 \\
    \hdashline[0.5pt/5pt]
    \begin{tabular}{@{}c@{}}Japan\\ Bunkyo (2.21)\end{tabular}               & 9.68  & 7.76  & 239.38 \\
    \hdashline[0.5pt/5pt]
    \begin{tabular}{@{}c@{}}Brazil\\ Sao Paulo (8.84)\end{tabular}           & 9.75  & 8.82  & 235.05 \\
    \hdashline[0.5pt/5pt]
    \begin{tabular}{@{}c@{}}CA, USA\\ Santa Clara (7.99)\end{tabular}        & 10.63 & 14.87 & 215.16\\
\end{tabular}
\vspace{0.05in}
\caption{ProtonVPN statistics.} 
\label{tab:vpn_summ}
\end{table}

Next, we compare the performance of the \emph{same} device at different locations. Since we do not have such testing condition, we emulate the presence of one device (J7DUO) at different locations via a VPN. We use a basic subscription to ProtonVPN~\cite{protonvpn} set up at the controller. Table~\ref{tab:vpn_summ} summarizes five locations we choose, along with network measurements from SpeedTest (upload and download bandwidth, latency). VPN vantage points are sorted by download bandwidth, with the South Africa node being the slowest and the California node being the fastest. Since the SpeedTest server is always within 10~km from each VPN node, the latency here reported is mostly representative of the network path between the vantage point and the VPN node.

Next, we leverage \webtool (see Section~\ref{sec:webtest}) to investigate the battery consumption of Chrome in comparison to a new privacy-preserving browser (Brave). We assume a simple workload where each browser is instrumented to sequentially load $10$ popular news websites. After a URL is entered, the automation script waits 6 seconds -- emulating a typical page load time (PLT) -- and then interact with the page by executing multiple ``scroll up'' and ``scroll down'' operations. 

\begin{figure}[t]
    \centering
    \psfig{figure=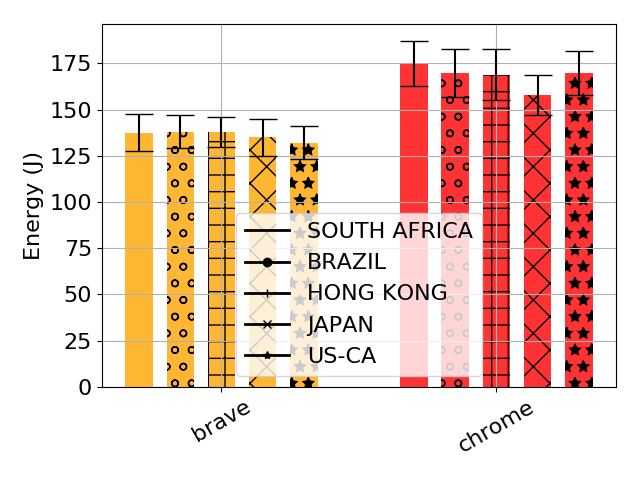, width=0.7\columnwidth}
    \caption{Brave and Chrome energy consumption measured through VPN tunnels.}
    \label{fig:eval:vpn}
\end{figure}

Figure~\ref{fig:eval:vpn} shows the average energy consumption (J) over 5 runs (standard deviation as errorbars) per VPN location and browser. The figure does not show significant differences among the battery measurements at different network location. For example, while the available bandwidth almost doubles between South Africa and California, the average discharge variation stays between standard deviation bounds. This is encouraging for experiments where \tool's distributed nature is a \emph{necessity} and its noise should be minimized. 

Figure~\ref{fig:eval:vpn} also shows an interesting trend when comparing Brave and Chrome when tested via the Japanese VPN node. In this case, Brave's energy consumption is in line with the other nodes, while Chrome's is minimized. This is due to a significant (20\%) drop in bandwidth usage by Chrome, due to a systematic reduction in the overall size of ads shown in Japan. This is an interesting result for experiments where \tool's distributed nature is a \emph{feature}.

\subsection{Usability Testing}
\label{sec:bench:usability}
The above analysis indicates that the extra (CPU) cost of device mirroring can invalidate the power measurements reading. Accordingly, our recommendation is to only leverage device mirroring when debugging an application over \tool, but then disable it during  the actual power measurements. This is not possible in case of usability testing where, by definition, a remote tester requires access to a device. 

``Action replay'' (see Section~\ref{sec:sys:action}) is a potential solution to this limitation. The intuition is that a usability test can be split in two parts. First, the tester performs the required test while action replay is used to record her actions. Then, the tester's actions are replayed without the extra cost associated with device mirroring. While this approach provides, in theory,  better accuracy, record and replay of human actions is challenging. 

\begin{figure}[t]
    \centering
    \psfig{figure=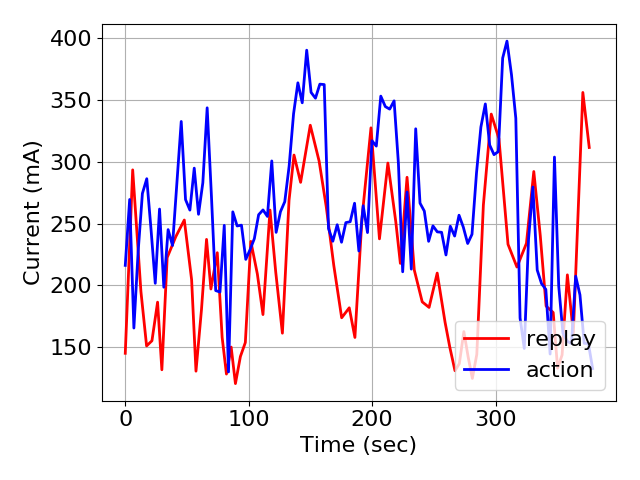, width=0.7\columnwidth}
    \caption{Evolution over time of current (mA) measured during ``action'' (one human interacting with 10 news websites in sequence) and ``replay'' (a bot reproducing human activity).}
    \label{fig:action-replay}
\end{figure}

Figure~\ref{fig:action-replay} shows current measured over time at SMJ337A when a human interacts with the news workload in Brave (\emph{action} curve), versus a bot generated by the action replay tool (\emph{replay} curve). The figure shows overall lower current usage in the replay case versus the action case, resulting in an overall lower energy consumption: 345 versus 399~J over 380 seconds. The figure also shows high correlation between the two curves, suggesting accurate replay. Nevertheless, the overall error for energy estimation is fairly low (about 10\%) and overall constant, similar to what suggested by~Figure~\ref{fig:relay} and now confirmed in a more challenging scenario. 

Record and replay of human actions is generally challenging. Not all applications behave equally between successive runs. For example, in our testing scenario, a webpage could suddenly slow down and the replay might fail to click on an article that was not loaded yet. This opens up a potential interesting research area of building ML-driven tools for app testing based on human input, as previously discussed in~\cite{almeida2018chimp}. This is currently out of the scope of this work, but pairing such research with device behavior monitoring is an interesting avenue for future work. 
\newcommand{\numsites}{{715}\xspace}

\section{The Web Power Monitor}
\label{sec:webtest}
There is an increasing interest in understanding the power drawn by modern websites and browsers~\cite{buiMOBICOM15,caoPOMAC17,thiagarajanWWW12}, especially on smartphones due to inherent battery constraints. Although different, these studies share a \emph{scalability} limitation: 1) they most target a single Android device, 2) they only test 100 websites or less accessed from a single location. This is because of the intrinsic limitations of the test-bed used which \tool aims at solving. In the following, we present the ``Web Power Monitor'' (\webtool), a \tool application which enable large scale measurements of energy consumption in the Web. \webtool currently powers a Web service~\cite{wpm-url} which offers such testing capabilities to the public, in real time. 

\begin{algorithm}[t!]
    \DontPrintSemicolon
    \KwIn{Device $device$, URLs to be tested $url\_list$, Browser $browser$, Number of repetitions $reps$, Power flag $power$, Visual flag $visual$, Automation $automation$}
    \KwOut{JSON file with performance metrics}

    \SetAlgoLined
    \SetKwComment{tcp}{$/\!\!/$ }{}
    \SetCommentSty{normal}
    \BlankLine
    
    {\it node\_status} $\leftarrow$ {\sc node\_setup}({\it power}, {\it visual})\\
    {\it device\_status} $\leftarrow$ {\sc device\_setup}({\it device})\\    
    \For(){$r \leftarrow 0$ \KwTo $reps$}{
        {\sc browser\_setup}({\it device}, {\it browser})\\
        {\sc run\_test}({\it device}, {\it browser}, {\it url\_list}, {\it automation})\\ 
    }
    {\it device\_status} $\leftarrow$ {\sc CLEANUP}({\it device})\\        
\caption{Pseudocode for \webtool's backend.}
\label{alg:power-meter}
\end{algorithm}

\subsection{Design and Implementation}
\label{sec:web:design}

\vspace{0.1in}
\noindent \textbf{Back-end} -- This is a Jenkins job whose goal is to report on the power consumed by a webpage when loaded on a \tool device via a desired browser. Algorithm~\ref{alg:power-meter} shows the pseudocode describing such job; for simplicity, we omit the part of the job that takes care to identify \emph{where} this test should run, \ie~either a specific device or a vantage point. As a first step (L1), the job prepares the vantage point by calling \texttt{NODE\_SETUP} (see~Section~\ref{sec:sys:api}) which activates power monitoring and device mirroring, if requested. Next, the device is prepared for the test using \texttt{DEVICE\_SETUP} (L2), also described in Section~\ref{sec:sys:api}. 

Next is  \texttt{BROWSER\_SETUP} (L3), where the browser is i) installed (if needed), ii) cleaned (cache and configuration files), and iii) freshly started, \eg~Chrome requires to go through an onboarding process when launched for the first time. This function is equivalent to what an experimenter would have to design for a local experiment, with the caveat that it has to be tested on a range of devices. \tool further simplifies this task for an experimenter via the ``action replay''  module (see Section~\ref{sec:sys:action}) which generates automation scripts from human input collected via \tool's browser interface. 

The next step is \texttt{RUN\_TEST} (L5) where a list of URLs ($url\_list$) is tested. The experimenter defines how URLs should be loaded, \eg~sequentially in a new tab. The experimenter also controls whether to perform a \emph{simple load}, \ie~load the page for a fixed amount of time, or interact with each page, \eg~scroll the page up and down multiple times for a certain duration. Finally, \texttt{CLEANUP} (see~Section~\ref{sec:sys:api})  is invoked to restore the node state (\eg~turn off the power meter) and, if needed, expose the collected data to the front-end.

\vspace{0.1in}
\noindent \textbf{Front-end} -- This is inspired by \texttt{webpagetest}~\cite{webpagetest}, a tool for measuring webpage load times.  Similarly, \webtool offers a simple Web interface where a visitor can choose the URL to test, the browser, the device (along with \emph{where} this device is located in the world) and which test to run. Power measurements are always collected, while visual access to the device needs to be explicitly requested. If so, the front-end alerts the user that this condition might impact the absolute value of the measurements, as discussed in Section~\ref{sec:bench}. 

Once the requested test is submitted, the user is presented with a page showing the progress of the experiment. If visual access was requested, as \texttt{DEVICE\_SETUP} is completed, an iframe is activated on the page to show the requested device in real time. At the end of \texttt{RUN\_TEST}, several plots are shown on screen such as CPU and current consumption during the test. We invite the interested reader to try out \webtool at~\cite{wpm-url}.

Front-end information (locations, devices, and browsers available) is maintained by the \texttt{REFRESH} job (see Section~\ref{sec:sys:api}) and then retrieved as a JSON file via an AJAX call. Similarly, the status of the experiment is pulled over time to switch between showing the remote device or results, when available.

\subsection{Results}
\label{sec:web:results}
We used \webtool to study the power consumption of Alexa's top 1,000 websites from 3 Android devices (NJ, IL, and UK) and one iPhone 7 (UK). We assume a simple load, where each page is loaded for up to 30 seconds. We load each page 3 times (with browser cache cleanup in between), and then report on the median for each metric.  We synchronize experiments at the different locations using a 2 minute fixed duration. The iOS test was not synchronized (one day delay) since only one testing device can be measured at a time per location.

To limit the scope (and duration) of this test, we only experiment with a single browser. We choose Brave, rather than Chrome as done in related work, since its ad-blocking feature offers a more consistent browsing experience location-wise (see Figure~\ref{fig:eval:vpn}) which in turn implies offering a similar workload to the different devices.  We invite the interest reader to leverage \webtool's online service to compare the energy consumption of different browsers.

\begin{table}[t]
\centering
\begin{tabular}{rccccc}
\hline
    {\bf Study}   & {\bf OSes} & {\bf Devices} & {\bf Websites} & {\bf Locations}\\
\hline
    \cite{thiagarajanWWW12}  & Android     & 1  & 25     & 1  \\
    \cite{caoPOMAC17}        & Android     & 1  & 80     & 1  \\
    \cite{buiMOBICOM15}      & Android     & 1* & 100    & 1  \\    
    \webtool                 & Android/iOS & 4  & 715    & 3  \\
\end{tabular}
\caption{\webtool's measurement study versus previous works.}
\label{tab:comp}
\end{table}

\begin{figure}[t]
    \centering
    \psfig{figure=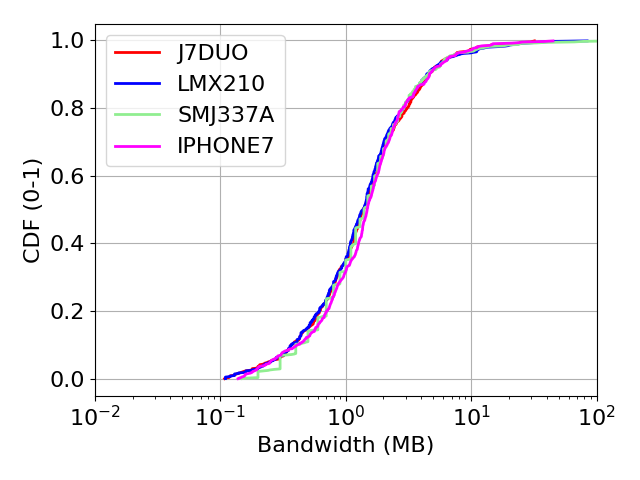, width=0.7\columnwidth}
    \caption{CDF of bandwidth consumption.}
    \label{fig:wpm-bdw}
\end{figure}

Table~\ref{tab:comp} summarizes the scale of this measurement in comparison with related work. We use Alexa's top 1,000 list from Nov. 2019 as an input. Out of the 1,000 websites we filtered 49 URLs potentially associated with adult websites---to avoid downloading and showing inappropriate content at our participating institutions---and 83 URLs associated with multiple top-level domains, \eg~\texttt{google.it} and \texttt{google.fr}. In the latter case, we kept the most popular domain according to Alexa, \ie~mostly the \texttt{.com} domain. Before the full experiment, we quickly tested the remainder URLs via a simple GET of the landing page. We find that 57 URLs had problems related with their certificate, 38 URLs timed out (30 seconds) and 58 URLs responded with some error code, 403 (Forbidden) and 503 (Service Unavailable) being the most popular errors. In the end, we are left with \numsites active URLs. 

We start by validating the assumption that Brave's ad-blocking helps in offering a similar workload to each device, irrespective of its location. Figure~\ref{fig:wpm-bdw} shows the CDF of the bandwidth consumed across websites when accessed from the three different device/locations. Overall, the figure confirms our assumption showing minimal difference between the four CDFs. Note that some variations are still possible because of, for instance, OS-specific differences, geo-located content, or consent forms that tend to be more prevalent in Europe due to the General Data Protection Regulation (GDPR)~\cite{gdpr}.

\begin{figure}[t]
    \centering
    \psfig{figure=\folder/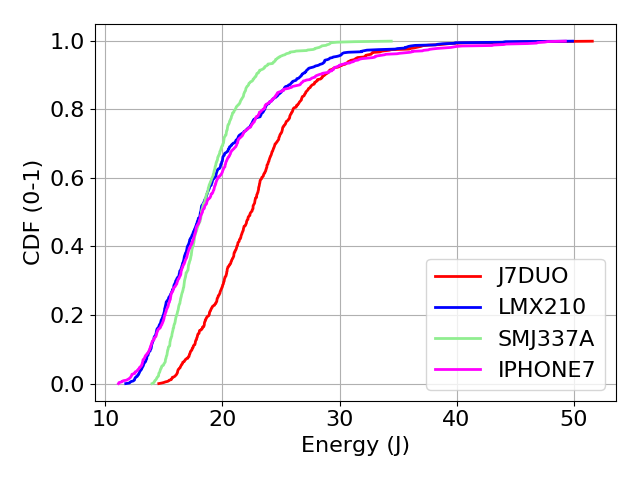, width=0.7\columnwidth}
    \caption{CDF of power consumption.}
    \label{fig:wpm-energy}
    \vspace{-0.2in}
\end{figure}

Figure~\ref{fig:wpm-energy} shows the CDF of the energy (J) consumed across websites, measured on each device. The J7DUO is by far the most power hungry device consuming, on average, 50\% more energy than both Android and the iOS devices --- a trend that has previously been observed during benchmarking (see Figure~\ref{fig:current-time}). A more similar trend is instead shared by the other devices, with most websites ($\sim$80\%) consuming between 10 and 20~J. The main differences can be observed in the tail of the distributions (10-20\%) where the results start to diverge. 

\begin{figure}[t]
    \centering
    \psfig{figure=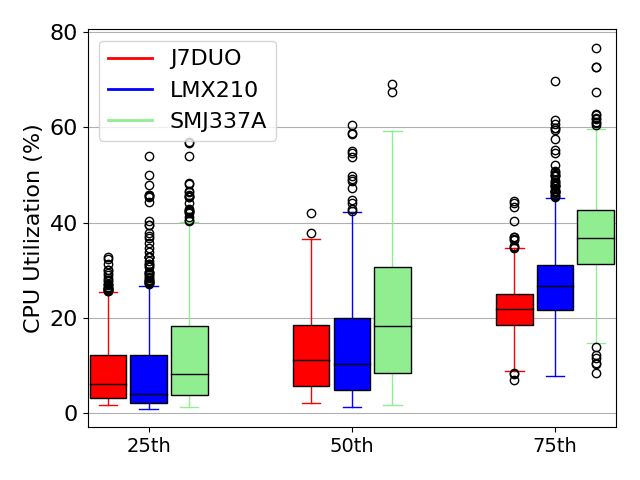, width=0.7\columnwidth}
    \caption{Boxplot with percentiles of CPU usage.}
    \label{fig:wpm-cpu}
\end{figure}

For completeness, we also analyze the CPU utilization during the test. We focus on Android only since, due to OS restrictions, it was not possible to obtain the CPU traces. We sample the CPU utilization every 3 seconds during each website load and report the 25th, 50th, and 75th percentile, respectively. Figure~\ref{fig:wpm-energy} summarizes this analysis as boxplots (across websites) for each percentile and device. An intuitive way to read this plot is to consider that low CPU values (25th percentile) refer to times before and after the CPU load, while the high CPU values (75th percentile) refer to times when the webpage was loaded. We observe that LMX210 and SMJ33, despite mounting similar hardware (see Table~\ref{tab:testbed}), exhibit different CPU usages, with SMJ33 spending overall more time at higher CPU utilization. The reason behind this are manifold, \eg~different OS versions and vendors. As expected, J7DUO suffers from less CPU pressure thank to its overall higher resources. 

To conclude, \tool allows to measure websites power consumption (and more) at unprecedented scale, in term of number of websites, devices, and network conditions. This opens up a new set of interesting research questions that we hope will appeal to the broader research community. 
\section{Related Work}
\label{sec:related}
Several commercial products --- such as AWS Device Farm~\cite{awsfarm}, Microsoft AppCenter~\cite{appcenter},  and Samsung Remote Test Labs~\cite{samsung} --- could leverage \tool's ideas to match our capabilities in a paid/centralized fashion. The same is true for startups like GreenSpector~\cite{greenspector} and Mobile Enerlytics~\cite{mobileenerlytics}, which offer software-based battery testing on few devices. 

In the research world, MONROE~\cite{alay2017experience} is the only measurement platform sharing some similarities with \tool. This is a platform for experimentation in operational mobile networks in Europe. MONROE currently has presence in 4 countries with 150 \emph{nodes}, which are ad-hoc hardware configurations~\cite{monroe_node} designed for cellular measurements. \tool is an orthogonal measurement platform to MONROE since it targets commercial devices (Android and iOS) and fine-grained battery measurements. The latter requires specific instrumentation (bulky power meters) that cannot be easily added to MONROE nodes, especially the mobile ones. Nevertheless, we are exploring BattOr~\cite{schulman2011phone}, a portable power meter, to enhance \tool with mobility support. 

Last but not least, \tool offers full access to test devices via a regular browser. This feature was inspired by~\cite{almeida2018chimp}, where the authors build a platform to grant access to an Android emulator via the browser to ``crowdsource'' human inputs for mobile apps. We leverage the same concept but also further extend it to actual devices and not emulators only. Further, remote access is just one tool in \tool's toolbox and not our main contribution. 
 
\section{Conclusion}
\label{sec:conclusion}
This paper has presented \tool, a collaborative platform for high-accuracy battery measurements where members contribute hardware resources (\eg~some phones and a power monitor) in exchange for access to the resources contributed by other platform members. To achieve this, we have built a complete prototyping suite which enables \emph{remote power testing} for both Android and iOS devices. By releasing our code and setup, we invite the community to join \tool, or at least we offer to eliminate the frustration associated with building ``yet-another'' home-grown performance measurement test-bed. 

\tool currently counts three vantage points, one in Europe and two in the US, hosting overall three Android devices and one iPhone 7. We evaluated \tool with respect to its accuracy of battery readings, system performance, and platform heterogeneity. We show that \tool's hardware and software have, for the most part, no impact on the accuracy of battery readings -- when compared with a ``local'' setup. This is not true when visual remote access to the device is required, \eg for \textit{usability} testing. However, \tool allows to ``record and replay'' usability tests which still offer accurate readings.  

Towards the end of the paper, we also demonstrated how to design and run large scale measurements via \tool. As an example, we have conducted, to the best of our knowledge,  the largest scale measurement study of energy consumption on the Web, encompassing Alexa's top 1,000 websites measured from four devices on both Android and iOS. We have further released a web application integrated with \tool which allows to measure the power consumption of a website, in real time. 

\section*{Acknowledgments}
This work was partially supported by the EPSRC Databox and DADA grants (EP/N028260/1, EP/R03351X/1).

\bibliographystyle{splncs04}
\small
\bibliography{biblio}

\begin{thebibliography}{10}
\providecommand{\url}[1]{\texttt{#1}}
\providecommand{\urlprefix}{URL }
\providecommand{\doi}[1]{https://doi.org/#1}

\bibitem{alay2017experience}
Alay, {\"O}., Lutu, A., Pe{\'o}n-Quir{\'o}s, M., Mancuso, V., Hirsch, T.,
  Evensen, K., Hansen, A., Alfredsson, S., et~al.: Experience: An open platform
  for experimentation with commercial mobile broadband networks. In: Proc. ACM
  MobiCom (2017)

\bibitem{almeida2018chimp}
Almeida, M., Bilal, M., Finamore, A., Leontiadis, I., Grunenberger, Y.,
  Varvello, M., Blackburn, J.: Chimp: Crowdsourcing human inputs for mobile
  phones. In: Proc. of WWW (2018)

\bibitem{mturk}
{Amazon Inc.}: {Amazon Mechanical Turk} (2022), \url{https://www.mturk.com/}

\bibitem{awsfarm}
{Amazon Inc.}: {AWS Device Farm} (2022),
  \url{https://aws.amazon.com/device-farm/}

\bibitem{route53}
{Amazon Inc.}: {Route 53 DNS} (2022), \url{https://aws.amazon.com/route53/}

\bibitem{appetize}
{Appetize}: {Run native mobile apps in your browser} (2022), \\
  \url{https://appetize.io/}

\bibitem{shareplay}
{Apple Inc.}: {SharePlay} (2021), \url{https://developer.apple.com/shareplay/}

\bibitem{AirPlay}
{Apple Inc.}: {How to AirPlay video and mirror your device's screen} (2022), \\
  \url{https://support.apple.com/HT204289}

\bibitem{blabtutorial}
{BatteryLab}: {Batterylab tutorial for new members} (2022), \\
  \url{https://batterylab.dev/tutorial/blab-tutorial.pdf}

\bibitem{wpm-url}
{BatteryLab}: {The Web power monitor} (2022), \\
  \url{https://batterylab.dev/test-website.html}

\bibitem{bluetooth_hid}
{Bluetooth SIG, Inc.}: {Human Interface Device (HID) Profile} (2022), \\
  \url{https://www.bluetooth.com/specifications/profiles-overview/}

\bibitem{bluez}
{BlueZ Project}: {BlueZ: Official Linux Bluetooth protocol stack} (2022), \\
  \url{http://www.bluez.org}

\bibitem{buiMOBICOM15}
Bui, D.H., Liu, Y., Kim, H., Shin, I., Zhao, F.: Rethinking energy-performance
  trade-off in mobile web page loading. In: Proc. ACM MobiCom (2015)

\bibitem{caoPOMAC17}
Cao, Y., Nejati, J., Wajahat, M., Balasubramanian, A., Gandhi, A.:
  Deconstructing the energy consumption of the mobile page load. Proc. of the
  ACM on Measurement and Analysis of Computing Systems  \textbf{1}(1),
  6:1--6:25 (Jun 2017)

\bibitem{chenSIGMETRICS15}
Chen, X., Ding, N., Jindal, A., Hu, Y.C., Gupta, M., Vannithamby, R.:
  Smartphone energy drain in the wild: Analysis and implications. In: Proc. ACM
  SIGMETRICS (2015)

\bibitem{gdpr}
{Data protection}: {Rules for the protection of personal data inside and
  outside the EU.} (2022),
  \url{https://ec.europa.eu/info/law/law-topic/data-protection_en}

\bibitem{RPiPlay}
{Florian Draschbacher}: {RPiPlay - An open-source AirPlay mirroring server for
  the Raspberry Pi.} (2022), \url{https://github.com/FD-/RPiPlay}

\bibitem{scrcpy}
{Genymobile}: {Display and control your Android device} (2022), \\
  \url{https://github.com/Genymobile/scrcpy}

\bibitem{adb}
{Google Inc.}: {Android Debug Bridge} (2022), \\
  \url{https://developer.android.com/studio/command-line/adb}

\bibitem{greenspector}
{Greenspector}: {Test in the cloud with real mobile devices.} (2022), \\
  \url{https://greenspector.com/en/}

\bibitem{ravenMOBICOM17}
Hwang, C., Pushp, S., Koh, C., Yoon, J., Liu, Y., Choi, S., Song, J.: Raven:
  Perception-aware optimization of power consumption for mobile games. In:
  Proc. ACM MobiCom (2017)

\bibitem{jenkins}
{Jenkins}: {The leading open source automation server} (2022),
  \url{https://jenkins.io/}

\bibitem{letsencrypt}
{Let's Encrypt}: {A a free, automated, and open Certificate Authority.} (2022),
  \\ \url{https://letsencrypt.org}

\bibitem{leung2016appforthat}
Leung, C., Ren, J., Choffnes, D., Wilson, C.: Should you use the app for that?:
  Comparing the privacy implications of app- and web-based online services. In:
  Proc. ACM IMC (2016)

\bibitem{our-code}
{Matteo Varvello, Kleomenis Katevas}: {BatteryLab Source Code.} (2022), \\
  \url{https://github.com/svarvel/batterylab}

\bibitem{appcenter}
{Microsoft, Visual Studio}: {App Center is mission control for apps.} (2022),
  \\ \url{https://appcenter.ms/sign-in}

\bibitem{mobileenerlytics}
{Mobile Enerlytics}: {The Leader In Automated App Testing Innovations To Reduce
  Battery Drain} (2022), \url{http://mobileenerlytics.com/}

\bibitem{monroe_node}
{MONROE - H2022-ICT-11-2014}: {Measuring Mobile Broadband Networks in Europe}
  (2022),
  \url{https://www.monroe-project.eu/wp-content/uploads/2017/12/Deliverable-D2.2-Node-Deployment.pdf}

\bibitem{monsoon}
{Monsoon Solutions Inc.}: {High voltage power monitor.} (2022), \\
  \url{https://www.msoon.com}

\bibitem{pymonsoon}
{Monsoon Solutions Inc.}: {Monsoon Power Monitor Python Library} (2022), \\
  \url{https://github.com/msoon/PyMonsoon}

\bibitem{novnc}
{noVNC}: {A VNC client JavaScript library as well as an application built on
  top of that library.} (2022), \url{https://novnc.com}

\bibitem{lucky2015wisec}
Onwuzurike, L., De~Cristofaro, E.: Danger is my middle name: Experimenting with
  ssl vulnerabilities in android apps. In: WiSec (2015)

\bibitem{protonvpn}
{ProtonVPN}: {High-speed Swiss VPN that safeguards your privacy} (2022), \\
  \url{https://protonvpn.com/}

\bibitem{rasbpi}
{Raspberry Pi}: {Raspberry Pi 3 Model B+} (2022),
  \url{https://www.raspberrypi.org/products/raspberry-pi-3-model-b-plus/}

\bibitem{jingjingrecon}
Ren, J., Rao, A., Lindorfer, M., Legout, A., Choffnes, D.: Recon: Revealing and
  controlling pii leaks in mobile network traffic. In: MobiSys (2016)

\bibitem{runthatapp}
{RunThatApp}: {Enjoy Mobile Apps In The Browser} (2022), \\
  \url{https://runthatapp.com}

\bibitem{samsung}
{Samsung}: { Remote Test Lab} (2022), \\
  \url{https://developer.samsung.com/remote-test-lab}

\bibitem{schulman2011phone}
Schulman, A., Schmid, T., Dutta, P., Spring, N.: Phone power monitoring with
  battor. In: Proc. ACM MobiCom (2011)

\bibitem{teamviewer}
{TeamViewer GmbH.}: {TeamViewer} (2022), \url{https://www.teamviewer.com/}

\bibitem{thiagarajanWWW12}
Thiagarajan, N., Aggarwal, G., Nicoara, A., Boneh, D., Singh, J.P.: Who killed
  my battery?: Analyzing mobile browser energy consumption. In: Proc. of WWW
  (2012)

\bibitem{tigervnc}
{TigerVNC}: {A high-performance, platform-neutral implementation of VNC
  (Virtual Network Computing).} (2022), \url{https://tigervnc.org}

\bibitem{usb_hid}
{USB Implementers' Forum}: {Universal Serial Bus HID Usage Tables} (2022), \\
  \url{https://www.usb.org/document-library/hid-usage-tables-112}

\bibitem{uhubctl}
{Vadim Mikhailov}: {uhubctl - USB hub per-port power control} (2022), \\
  \url{https://github.com/mvp/uhubctl}

\bibitem{varvello2019batterylab}
Varvello, M., Katevas, K., Plesa, M., Haddadi, H., Livshits, B.: Batterylab, a
  distributed power monitoring platform for mobile devices. In: HotNets (2019)

\bibitem{webpagetest}
{Webpagetest}: {Test website performance} (2022),
  \url{https://www.webpagetest.org/}

\bibitem{elan}
Wittenburg, P., Brugman, H., Russel, A., Klassmann, A., Sloetjes, H.: Elan: a
  professional framework for multimodality research. In: LREC. vol.~2006 (2006)

\end{thebibliography}

\end{document}